\documentclass[aps,nofootinbib,showpacs,twocolumn]{revtex4}
\usepackage[CJKbookmarks, pdftex, bookmarksnumbered, bookmarksopen, colorlinks, citecolor=blue, linkcolor=blue]{hyperref}
\usepackage[english]{babel}
\usepackage{amstext,amsmath,amssymb,amsfonts,bbm}
\usepackage[latin1]{inputenc}
\usepackage{graphicx}
\usepackage{epstopdf}
\usepackage{amsthm}
\usepackage{tocvsec2}
\usepackage{enumerate}
\usepackage{subfigure}
\usepackage{color}
\usepackage{tikz}
\usepackage[squaren,Gray]{SIunits}

\newcommand{\be}{\nopagebreak[3]\begin{equation}}
\newcommand{\ee}{\end{equation}}

\newcommand{\bee}{\nopagebreak[3]\begin{equation*}}
\newcommand{\eee}{\end{equation*}}

\newcommand{\ba}{\nopagebreak[3]\begin{eqnarray}}
\newcommand{\ea}{\end{eqnarray}}
\DeclareFontFamily{U}{rsfs}{}         
\DeclareFontShape{U}{rsfs}{m}{n}{<5> rsfs5 <6><7> rsfs7          %
  <8><9><10><10.95><12><14.4><17.28><20.74><24.88> rsfs10}{}     %
\DeclareMathAlphabet{\mathfs}{U}{rsfs}{m}{n}                     %
\newcommand{\mfs}[1]{\mathfs {#1}}                               %
\newcommand{\n}{{\nonumber}}

\newcommand{\sO}{{\mfs O}}

\begin{document}

\title{Dark energy from quantum gravity discreteness}

\author{Alejandro Perez}
\affiliation{{Aix Marseille Univ, Universit\'e de Toulon, CNRS, CPT, Marseille, France}}

\author{Daniel Sudarsky}
\affiliation{Instituto de Ciencias Nucleares, Universidad Nacional Aut\'onoma de M\'exico, M\'exico D.F. 04510, M\'exico.\\
Department of Philosophy, New York University,  New York, NY 10003. }

\date{\today}
\begin{abstract}

  We argue that discreteness at the Planck scale (naturally expected to arise from quantum gravity) 
might  manifest  in  the form  of  minute violations of energy-momentum conservation of the matter degrees of freedom 
when described in terms of (idealized) smooth fields on a smooth spacetime.  In the context of applications to cosmology such 
`energy   diffusion' from the low energy matter degrees of freedom to the discrete structures 
underlying  spacetime leads to the emergence  of 
 an effective  dark energy  term  in  Einstein's equations. We estimate this effect using a (relational) hypothesis about the materialization of discreteness in quantum gravity which is motivated by the  strict   observational constraints supporting the  validity of Lorentz invariance at low energies. Arguments based on a simple dimensional analysis lead to an estimate of an effective cosmological constant agreeing in order of magnitude with its observed value. If correct, this would constitute remarkable empirical evidence for a Planckian granular aspect of spacetime. 
	\end{abstract}

\maketitle

The  discovery that the universe is undergoing an accelerated expansion \cite{Riess:1998cb, Perlmutter:1998np} 
is the source of one of the greatest puzzles of our present understanding of cosmology which goes under the name of the {\em dark energy} problem. While the assumption of the presence of a cosmological constant $\Lambda$ remains the most successful phenomenological model,  naive theoretical reasoning predicts a value for $\Lambda$ that is either $120$ orders of magnitude   to big,  or  is strictly vanishing when a protective symmetry principle is at play \cite{Weinberg:1988cp}. It would  be  desirable to have a   concrete  fundamental calculation  leading clearly  to $\Lambda_\text{obs} \approx 1.19~10^{-52}~\meter^{-2}$, the value indicated by observations \cite{Adam:2015rua}. 

{A recent work  \cite{Josset:2016vrq}  proposed} a framework where violations of energy momentum conservation
produce a dark energy contribution. The key result of {that  work was to characterize }
the effective framework where violations of energy conservation are made compatible with general relativity.  
As an illustration we applied it to two models,   {previously considered in the   literature,  that propose such violations}. 
However, none of these two {could  be  taken as  truly  realistic.  On the one hand, the cosmological time at which the effects 
would start was not intrinsically defined by the models, and, on the other hand, the strength of the violations of energy conservation 
were encoded in a phenomenological   adjustable parameter with no explicit link to fundamental constants. Therefore,  while these examples were  illustrative of  the idea that small violations
 can accumulate and contribute non negligibly to  $\Lambda$,  they could not be used to predict its value.}

In this paper we bridge this gap by proposing a mechanism  to  generate  $\Lambda $  and  the  quantitative estimates  based  entirely  on  known fundamental features of the physics involved. The origin of the cosmological term, we suggest, is to be found in the
microscopic structure of spacetime and its interaction with matter. We  will work under the 
hypothesis that discreteness of geometry and Lorentz invariance at low energies are fundamental aspects of  quantum gravity. Based on these two fundamental features we propose a phenomenological model for quantum-gravity-induced violations of energy conservation   depending only  on the  fundamental constants $G,c, \hbar$ and  a few parameters of the standard model (SM).
We show that our simple proposal resolves the two limitations of the 
previous examples and predicts a contribution to the cosmological constant of the correct order of magnitude.

One of the most important constraints on the form of quantum discreteness at Planck scales comes from the observed
 validity of Lorentz invariance at QFT scales. {As  shown in \cite{Collins:2004bp, Collins:2006bw}  this rules out the simple atomistic view of a spacetime foam selecting a preferred `rest-frame' at the Planck scale. This result, which  severely constrains phenomenological ideas, is corroborated } by a large collection of empirical evidence  \cite{Mattingly:2005re}.   A  more subtle theoretical characterization of space-time  discreteness at Planck's scale is necessary.  
 
We {think } that the key for understanding Planckian discreteness lies in the relational nature of physics partly  uncovered by Einstein's theory of gravity \cite{reichenbach2012philosophy}.
In general relativity, geometry  can only be probed by  the matter  degrees of freedom. 
The metric has a clear  physical 
meaning only when rulers and clocks are introduced. 
 More precisely, the construction of observables  (diffeomorphism invariant quantities)  requires the use of  relational notions involving  a mixture of  geometric and 
matter  degrees of freedom. The difficulty of actually defining such quantities is,  in fact,  one of the most severe  technical problems in formal approaches to quantum gravity.    In our view such relational perspective is essential for understanding discreteness at the Planck scale. 
We are {thus } rejecting the notion of  a spacetime foam acting as an  empty  arena  where  matter,  if there  placed,   would  reveal  its  preexisting features. Quantum discreteness should arise primarily via the interactions  of  gravity  with those other degrees of freedom,  which  by  their nature,  are able to select a  preferential  rest-frame  where  the fundamental scale $\ell_p$ acquires an invariant meaning. In other words, within  the  relational approach  we  are advocating,  it  is clear that, in  order  to be directly sensitive to the discreteness scale $\ell_p$, the probing degrees of freedom must themselves carry their intrinsic scale. Thus massless (scale-invariant) fields are ruled out as leading probes of discreteness simply because they cannot  be associated  with any local notion of rest  frame, {and  thus,  of a fundamental length scale}.    
This argument {identifies } massive fields as the natural candidates for probes of {spacetime}  discreteness. Such  discreteness  must be  thus thought as becoming relevant, or  as `awaken',  by the interactions of gravity with such scale-invariance-breaking fields.
 The immediate possibility arising from such considerations  (and  framed  in a phenomenological perspective) is that  low energy quantum field theoretical excitations of massive  fields could interact with the underlying  quantum gravity  microstructure and exchange `energy' 
with it \footnote {Some  ideas  with  similar  conceptual underpinning have been explored in the  context of  laboratory searches for quantum gravity phenomenology \cite{Corichi:2005fw, Bonder:2007bj, Aguilar:2012ju}. For a discussion of the implications for the information problem in black hole evaporation  see \cite{Perez:2014xca, Perez:2017cmj}.}.

In order to study the phenomenological implications of these ideas one needs a  `mean field' or macroscopic description
of the quantity parametrizing the phenomenon.  An obvious choice is the trace of the energy-momentum tensor $\mathbf T$---which for a fluid in 
thermal equilibrium is simply given by ${\mathbf T}=-\rho+3P$---which signals the breaking of conformal invariance, and hence, 
the presence of massive degrees of freedom. Via Einstein's equations $\mathbf T$ is related to the scalar curvature ${\mathbf R}=-8\pi G {\mathbf T}$.
Therefore, the presence of massive  fields (suitable probes of discreteness according to our rationale) is geometrically captured by a non trivial $\mathbf R$. 

The effect on the propagation of massive fields must be realized in a deviation from the geodesic motion of free particles due to a `friction-like' force encoding the noisy interaction with the underlying spacetime granularity. As argued in the previous paragraphs, the force must be proportional to $\mathbf R$. In addition, the force should depend on the mass $m$, the 4-velocity $u^{\mu}$, the spin $s^{\mu}$ of the classical particle (the only spacetime related intrinsic features defining a particle), and a time-like unit  vector  $\xi^{\mu}$ specifying the local frame  defined  by the matter that curves spacetime. For instance, in cosmology $\xi=\partial_t$
is naturally associated with the time-arrow of the  co-moving cosmic  fluid.  In addition, and according to our preceding  arguments, the force should be proportional to the particle's mass, endowing it with a  characteristic length scale: the Compton wave-length. Dimensional analysis gives an essentially  unique expression   which  is   compatible with the above requirements \footnote{Higher curvature corrections could be added, but these are highly suppressed by the Planck scale and are thus negligible for the central point of this letter. A term proportional to $\epsilon^{\nu}_{\ \mu\gamma\sigma} \xi^\mu s^\gamma u^\sigma$ is also allowed but does not affect the results.},  
\ba\label{modimodi}
u^{\mu}\nabla_{\mu}  u^{\nu}&=&\alpha \frac{m}{m^2_p}\, {\rm sign}(s\cdot \xi) { \mathbf R}\, {s^{\nu}},   
\ea
where $\alpha> 0$ is a dimensionless coupling\footnote{\label{cinco} 
There is a remarkable formal similarity of equation \eqref{modimodi} with  others arising in well understood situations.   We have  the Mathisson-Papapetrou-Dixon equations  \cite{Papapetrou:1951pa}  describing the dynamics of idealized extended objects in GR,  
  \be
\label{Papap1} 
u^{\nu} \nabla_{\nu} P_{\mu} 
=
-\frac{1}{2}  {\mathbf R}_{\mu\nu \rho  \sigma} u^{\nu} S^{\rho  \sigma},\n
\ee
where $u^{\mu}$ represents the 4-velocity of the object, $P^{\mu}$ its 4-momentum, $S^{\rho  \sigma}$ its spin and 
$R_{\mu\nu \rho  \sigma}$ is the Riemann tensor.  
Moreover,  we  note that  the characterization of WKB-trajectories of the Dirac theory on a pseudo-Riemannian geometry \cite{Audretsch:1981xn}, to lowest order in $\hbar$, is given by
\be\label{wkb}\n
u^\nu \nabla_\nu (m u_\mu) =-\frac{1}{2} \tilde {\mathbf R}_{\mu\nu \rho  \sigma} u^\nu \langle S^{\rho\sigma}\rangle +\sO(\hbar^2).
\ee
The previous is equivalent to (\ref{modimodi}) if one considers   an effective $\tilde {\mathbf R}_{\mu\nu \rho  \sigma}\propto  {m^2}/{m_p^2}\,  {\rm sign}(s\cdot\xi) {\mathbf R} \epsilon_{\mu\nu \rho  \sigma}$ taken to  encode a pure torsion-related  structure as $\tilde {\mathbf R}_{[\mu\nu \rho]  \sigma}\not=0$ (from the first Bianchi identities). 
}.

The factor ${\rm sign}(s\cdot \xi)$  makes the force genuinely friction-like.
  This is apparent when one considers  the change of the mechanical energy of the particle  $E\equiv -m u^{\nu}\xi_{\nu}$ (defined in the frame defined  by $\xi^{\mu}$) along the particles world-line, namely
\ba\label{conserv} \dot E&\equiv& -m u^{\mu}\nabla_{\mu}(u^{\nu}\xi_{\nu})\n \\ &=&-\alpha  
\frac{m^2 }{m^2_p} |(s\cdot \xi)| {\mathbf R}-m u^{\mu}u^{\nu}\nabla_{(\mu}\xi_{\nu)}.\ea
The last term  in \eqref{conserv} encodes the  standard  change of $E$  associated to the  non-Killing  character of  $\xi^{\mu}$.
The first term on the right  encodes the friction that damps out any motion with respect to $\xi^\mu$.  Energy is lost into the fundamental granularity until $u^{\mu}=\xi^{\mu}$ and the particle is at rest with the cosmological fluid, and thus $\dot E=0$.

The simplest dynamics for the spin that is consistent with  (\ref{modimodi}), the conservation of $s\cdot s$, and $s\cdot p=0$ is \be\label{spinny} u^{\mu}\nabla_{\mu} s^{\nu}=\alpha \frac{m}{m^2_p}\, {\rm sign}(s\cdot \xi) { \mathbf R}\, (s\cdot s)\, u^{\nu}.\ee
This is only a minimalistic solution, other terms can be added.
We will investigate these aspects elsewhere as they might be important for 
phenomenology; however, they do not affect the main point in this letter.

In this respect, it is also important to point out that the violations of the equivalence principle and Lorentz invariance implied by \eqref{modimodi} and (\ref{spinny}) can be readily  checked  not to  be in conflict with well known observational bounds by many orders of magnitude \cite{Kostelecky:2008ts} for $\alpha \sim  O(1)$. A simple indication comes  from  comparison  of  the value of $\mathbf R$ at the electro weak (EW) transition in  cosmology (a regime where our effects will be important) to that associated  with, say, the gravitational effect of a  piece of lead: this gives $\frac{{\mathbf R}_{lead}} {{\mathbf R}_{EW}}  \sim 10^{-24}$.

Coming back to the main argument, the diffusion of energy for a single particle, induced  by (\ref{modimodi}), implies  the lack of energy-momentum conservation for a fluid constituted by an ensemble of such particles (we will compute this  below). However, violations of energy-momentum conservation are incompatible with general covariance and hence with the standard general relativity description of gravity. Fortunately, there is a simple relaxation of general covariance (originally studied by Einstein) from full coordinate invariance  down to spacetime volume preserving coordinate transformations. Such modification---which we 
only take as an effective low energy description of a (in a suitable sense) general covariant fundamental physics---is called unimodular gravity (UG), 
and its field equations are just the trace-free part of the standard Einstein's equations
 \begin{equation}\label{TraceFreeEinsteinEquation}
	{\mathbf R}_{\mu\nu} - \frac{1}{4} {\mathbf R} g_{\mu\nu} = {8 \pi G} \left( {\mathbf T}_{\mu\nu} - \frac{1}{4} {\mathbf T}g_{\mu\nu} \right).
\end{equation}
Defining $J_\mu\equiv (8\pi G) \nabla^\nu T_{\nu\mu}$, assuming UG integrability $dJ=0$ , and using Bianchi identities, one obtains \cite{Josset:2016vrq}
\begin{equation}\label{TraceFreeEinsteinEquation2}
{\mathbf R}_{\mu\nu} - \frac{1}{2} {\mathbf R} g_{\mu\nu} +\underbrace{\left[\Lambda_{ 0} +\int_{\ell} J\right] }_{\Lambda}g_{\mu\nu}= 
{8 \pi G}  {\mathbf T}_{\mu\nu} ,
\end{equation}
where $\Lambda_0$ is a constant of integration, and $\ell$ is a one-dimensional path from some reference event. Thus, the energy-violation current $J$ is the source of a term in Einsteins equations satisfying the dark energy equation of state. An additional, often alluded  feature of  UG  is that the vacuum energy does not gravitate  \cite{Weinberg:1988cp, Ellis:2010uc, Ellis:2013eqs}.

Now we compute $J_\nu\equiv 8\pi G \nabla^\mu T_{\mu\nu}$ as implied by \eqref{modimodi}. 
For a particle species $i$ (the  interactions  between different species are  neglected here   as  their effect lead only to  very  small  corrections) one has ${\mathbf T}_{\mu\nu}^i$ \footnote{There  is  a   subtle   point that  ought to  be noted  here:  this part of the   calculation  is carried  out  by  considering   a space-time  region  small enough to be covered  by Riemann normal  coordinates  (i.e.  a local inertial frame) in  such a  way that the standard  effects of curvature  can be  neglected. The region is however large in comparison with the Planck scale so that the energy diffusion effects, the non standard influence of $\mathbf R$ in our model,  are encoded in the friction force underlying (\ref{modimodi}).}
\be\label{titi}
{\mathbf T}^i_{\mu\nu}(x)\equiv \int p_{\mu} p_{\nu} \, f^i(x, p, s_r) {\rm D}p {\rm D}s_r,
\ee
where $f^i(x, p,s_r)$ encodes the particle distribution in phase space with $s_r$ denoting the value of the spin of the particle in its rest frame, ${\rm D}p=\delta(p^2+m^2)dp^4$, and ${\rm D}s_r$ is the standard measure on the sphere of the spin directions. Simple kinetic theory allows to express $\nabla^\mu{\mathbf T^i}_{\mu\nu}$ as (see equation 2.113 in \cite{rezzolla2013relativistic}) 
\ba\label{calcul}
\frac{\nabla^\mu{\mathbf T}^i_{\mu\nu}}{{\mathbf T}^i}
&=& -\frac{\int m_i F_{\nu} f^i(x, p,s_r)  {\rm D}p {\rm D}s_r }{m_i^2 \int f^i(x, p, s_r) {\rm D}p {\rm D}s_r}   \\ 
&=& -\alpha \frac{m_i}{m_p^2} {\mathbf R} \frac{ \int  \left[\frac{ s_{\nu} s_0}{ |s_0|}\right] f^i(x, p,s_r)  {\rm D}p {\rm D}s_r }{\int f^i(x, p,s_r)  {\rm D}p {\rm D}s_r } \n
\ea
where $0$-components refer to the time direction $\xi^\mu$.
Assuming thermal equilibrium at temperature $T$, and ignoring the negligible additional effects of the force on the distribution, we have $f^i(x,p,s_r)=f^i_{T}(p)$ where the later is the standard Boltzmann distribution.  

Isotropy of the equilibrium configuration implies that only the $0$-component of \eqref{calcul} is non trivial. 
Then the result follows first 
 from the fact that 
 \be
 \int |s_0| {\rm D}s_r= \frac{2\pi {\mathbf p}|s|}{m} \int  |\cos(\theta)| \sin(\theta) d\theta  = \frac{2\pi \mathbf{p}|s|}{m},\n
 \ee
 where ${\mathbf p}^2\equiv \vec p\cdot \vec p$, and the factor ${\mathbf p}/m$ comes from the boost relating the comoving frame to the rest frame of the particle. The next step is
 \be\label{97}
\frac{ \int  \left[ \frac{2\pi {\mathbf p}|s|}{m}\right] f_{T}(p) {\rm D}p  }{ \int f_{T}(p) {\rm D}p }={4 \pi |s|} \frac{T}{m}\left[1+\sO\left(\log\left(\frac{m}{T}\right)\frac{m^2}{{T^2}}\right)\right].\n
\ee
Therefore, in the relativistic regime $T\gg m$ one has 
 \ba\label{jojo}
J_\nu&\equiv& (8\pi G)\, \nabla^\mu {\mathbf T}_{\mu\nu}= {4 \pi \alpha } \frac{T  }{m_p^2}  {\mathbf R} \left[8 \pi G\sum_i |s_i| {\mathbf T}^i \right] \xi_\nu,\n \\
&\approx&  {2 \pi \alpha \hbar} \frac{T  }{m_p^2}  {\mathbf R}^2 dt_\nu
\ea
where in the last line we write an approximation valid for the case where a single $|s|=\hbar/2$ fermion species dominates.
This approximation will be useful in the application of the formula to cosmology.

We now focus  on the effects of (\ref{jojo}) in the  dynamics of the early universe when its macroscopic geometry is well approximated by the flat  FLRW metric
\be
ds^2=-dt^2+a(t)^2 d\vec x^2,
\ee
and where the local frame $\xi=\partial_t$ is  identified  with co-moving observers. As only massive  particles  with spin are  subjected  to the 
frictional force  \eqref{modimodi},  the diffusion mechanism in cosmology starts when such particles first appeared. According to the standard model---whose validity is assumed from the end of inflation---this corresponds to the electro-weak (EW) transition time. 
We further assume that a protective symmetry enforces  $\Lambda_0=0$ (see for instance \cite{Hawking:1984hk,Coleman:1988tj}).

\begin{figure}[h]
\centerline{\hspace{0.5cm} \(
\begin{array}{c}
\includegraphics[height=5.cm]{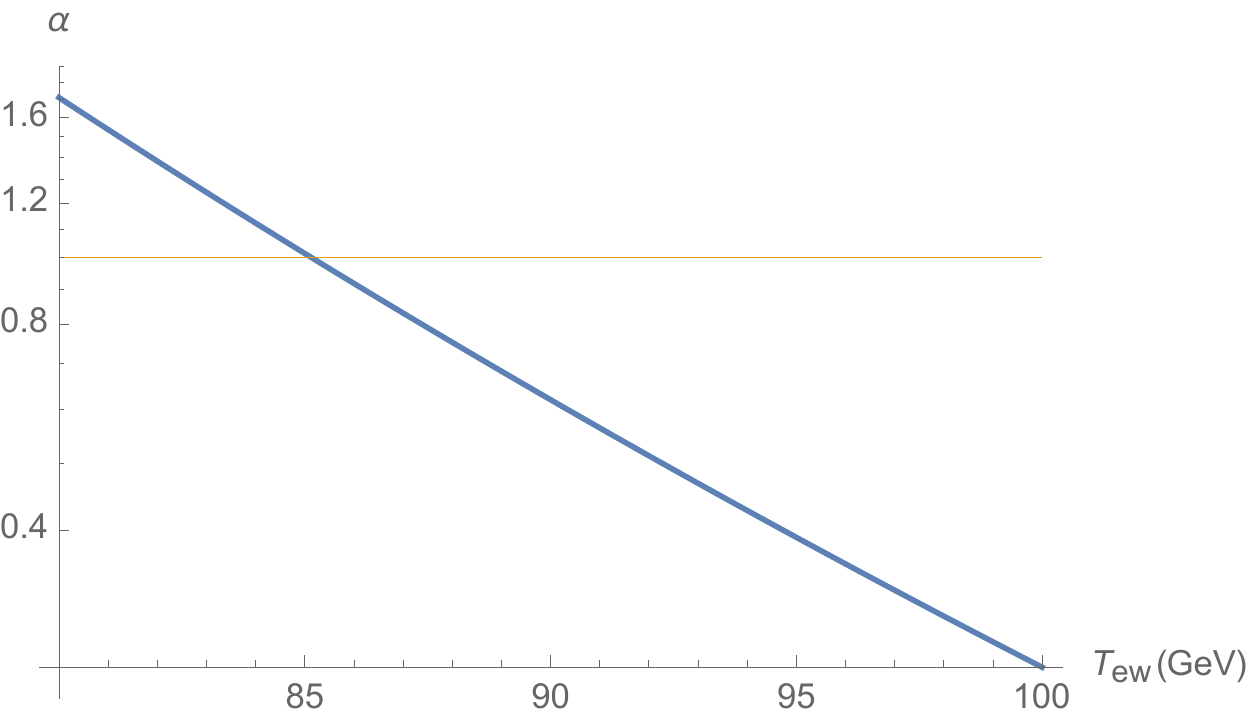} 
\end{array}\) } \caption{The value of the phenomenological parameter $\alpha$, see eq. (\ref{modimodi}), that fits the observed value of $\Lambda_{obs}$ as a function of the EW transition scale $T_{\rm ew}$ in $\rm GeV$. The  contributions  from  the massive gauge bosons  of the standard model have been included. 
}
\label{figu}
\end{figure}

We are {now } ready to estimate the effective cosmological constant predicted by our model. Using \eqref{TraceFreeEinsteinEquation2}, and (\ref{jojo}) one gets
\be\label{LL}
\Lambda=\frac{2\pi \alpha \hbar}{m_p^2} \int\limits_{t_{\rm ew}}^{t_0} \left[{8\pi G} ( \rho-3 P)\right]^2  T \, dt,
\ee	
{with $t_0$  the present time. 
It is convenient to change the integration variable in (\ref{LL})   from  co-moving time $t$ to  temperature $T$ given the essentially   direct  relation  between  the two quantities. 
During the relevant period, of radiation domination, the  matter fields are assumed to be  in thermal equilibrium.}  The density of the universe 
is then given by, 
$\rho={\pi^2 g_*T^4}/{(30\hbar^3)}$
where $g_*\approx 100$ is the effective degeneracy factor for the temperatures of interest \cite{Peter:1208401}.
Taking into account that  temperature scales like $a^{-1}$, using Friedman equation, and $H(a)=\dot a/a$,  one gets,
\be\label{dt}
\frac{dT}{T}=-\frac{da}{a}
=-\underbrace{\sqrt{\frac{8\pi G}{3} \frac{\pi^2 g_* T^4}{30 \hbar^3}} }_{H(a)}dt.
\ee We  will  now focus   just on the   leading  contributions.
In the ultra-relativistic regime standard thermodynamics leads to the expression
\be\label{r}
 \rho-3P\approx
\frac{m_t^2 T^2}{2 \hbar^3}  ,
\ee
where $m_t$ is the top mass. Replacing the leading term in \eqref{r} and (\ref{dt}) into \eqref{LL} one gets 
\ba\label{wwww}
\Lambda&\approx&
{{16 \alpha}   \sqrt{\frac{5 \pi^3}{g_*}}}\ \frac{  m_t^4 {T_{\rm ew}}^3}{m_p^5\hbar^2  } \ \epsilon(T_{\rm ew}),\ea 
where
\be
\epsilon(T_{\rm ew})=-\frac{3}{T^3_{\rm ew}} \int\limits_{T_{\rm ew}}^{T_{\rm end}} \left(1-\frac{T^2}{T_{\rm ew}^2}\right)^2 T^2 dT
\ee
is a dimensionless correction factor that takes into account the temperature dependence of the quark mass during the EW-transition, namely $m^2_t(T)=m_t^2 (1-T^2/T^2_{\rm ew})$. The end temperature $T_{\rm end}$ is the one satisfying $2m_t(T_{\rm end})=T_{\rm end}$ when the top quark's  abundance  decreases  dramatically. The contribution of other fields in the standard model, as  well as  those  tied to  simple  dark matter  models such as  WIMPS  will not  affect the order of magnitude of the estimate \footnote{\label{GB} Massive gauge bosons do not change the order of magnitude estimate, as ${m_Z}/{m_t}\approx 1/2$ and ${g_{ZW^{\pm}}}/{g_{t\bar t}}=3/4$. In (\ref{wwww}) this leads to a factor $(3/4)^2 (1/2)^4$ times $2$ as the spin of the bosons is twice that of the fermions, i.e. their   contributions is   about  $7 \%$ of that  coming from top-quark. From (\ref{jojo}) one can work out the precise corrections which are included in Figure \ref{figu}.}.  We note that aside from the correction factor, $\epsilon(T)\approx 10^{-3}$---$10^{-4}$ in the range of interest, equation 
(\ref{wwww}) could have been guessed from dimensional analysis.
After substitution of the different quantities involved and taking for example $T_{\rm ew}\approx 100 \ {\rm GeV }$ \cite{Carrington:1991hz, Morrissey:2012db}, and adding the gauge boson contributions (not included in \eqref{wwww}) we find 
\be  \Lambda \approx 4\,  \alpha\, \Lambda_{\rm obs}\ee
where ${\Lambda_{\rm obs}}$ is the observed value of the cosmological constant. For other values of $T_{\rm ew}$  see Figure \ref{figu} where we plot the value of the dimensionless coupling $\alpha$ needed to fit the observed values as a function of $T_{\rm ew}$.  These results are an order of magnitude estimate;  a refined calculation   would  require   detailed considerations  of the  dynamics of the electro-weak transition. However, such  details are  not  expected  to  modify  our  result in essential  ways. 

We believe that our proposal has important implications of various  types.
At the theoretical level it provides a novel view that could reconcile Planckian discreteness and Lorentz invariance and gives possibly valuable insights guiding  the search for  a theory  of  quantum gravity.  At the  empirical level  our analysis opens a new  path  for  searches of new  physical  manifestations of the   gravity/quantum  interface.  
     
 Concerning the later,   we  note that one might  use \eqref{jojo} to estimate the amount of energy loss in local experiments.  Presently (neglecting  the cosmic expansion), we find
$\dot \rho \approx- \alpha  ({\rho}/{\rho_{\rm water}})^2 10^{-70} {g}/({cm^{3} s})$
where $\rho_{\rm water}$ is the density of sea water. 
The amount of energy produced is maximal at the EW transition when the density of the universe  $\rho(T_{\rm ew})\approx  \  10^{25} g/cm^3$,  and corresponds to a relative changeof energy density in a Hubble time of $\Delta\rho/\rho\approx \alpha \, 10^{-51}$ .  Such  a minuscule   level  of  energy loss cannot have  significant  effects   on  the matter dynamics, and thus would be very hard, but not impossible to detect. 
 Nevertheless,  we have seen that such small energy losses can explain the  observed   late time acceleration of the expansion rate of our universe.

Finally, as  the model links $\rho$ and its evolution  with the present value of the cosmological constant, and 
$\rho$ directly enters in the computation of the structure formation  leading  to   galaxies,  stars  and eventually humans, this framework  opens,  in principle,  a path  that might help   in addressing the longly debated `coincidence problem' \cite{Peter:1208401}.

\section*{Acknowledgments}
We acknowledge fruitful interactions with James Bjorken. We are also grateful for the constructive interaction with the anonymous referees.
DS  acknowledges partial financial support from DGAPA-UNAM project IG100316 and  by CONACyT project 101712, as  well as the  sabbatical  fellowships  from  CO-MEX-US (Fulbright-Garcia Robles)  and  from  DGAPA-UNAM  (PASPA).  AP acknowledges the OCEVU Labex (ANR-11-LABX-0060) and the A*MIDEX project (ANR-11-IDEX-0001-02) funded by the ``Investissements d'Avenir" French government program managed by the ANR.

\end{document}